\documentclass[twocolumn]{aastex631}

\shorttitle{Glass-based particle loads}
\shortauthors{Liao et al.}

\graphicspath{{./}{figures/}}

\newcommand{\vect}[1]{\mathbf{#1}}

\begin{document}

\title{Particle loads for cosmological simulations with equal-mass dark matter and baryonic particles}

\author[0000-0001-7075-6098]{Shihong Liao}
\affiliation{Key Laboratory for Computational Astrophysics, National Astronomical Observatories, Chinese Academy of Sciences, 20A Datun Road, Chaoyang District, Beijing 100101, China}
\affiliation{School of Astronomy and Space Science, University of Chinese Academy of Sciences, Beijing 100049, China}

\author[0009-0005-8855-0748]{Yizhou Liu}
\affiliation{Institute for Frontiers in Astronomy and Astrophysics, Beijing Normal University, Beijing 102206, China}

\author[0000-0002-1665-5138]{Haonan Zheng}
\affiliation{Kavli Institute for Astronomy and Astrophysics, Peking University, Beijing 100871, China}

\author[0000-0002-1318-4828]{Ming Li}
\affiliation{Key Laboratory for Computational Astrophysics, National Astronomical Observatories, Chinese Academy of Sciences, 20A Datun Road, Chaoyang District, Beijing 100101, China}
\affiliation{School of Astronomy and Space Science, University of Chinese Academy of Sciences, Beijing 100049, China}

\author[0000-0002-9937-2351]{Jie Wang}
\affiliation{Key Laboratory for Computational Astrophysics, National Astronomical Observatories, Chinese Academy of Sciences, 20A Datun Road, Chaoyang District, Beijing 100101, China}
\affiliation{School of Astronomy and Space Science, University of Chinese Academy of Sciences, Beijing 100049, China}

\author[0009-0006-3885-9728]{Liang Gao}
\affiliation{Institute for Frontiers in Astronomy and Astrophysics, Beijing Normal University, Beijing 102206, China}
\affiliation{School of Physics and Microelectronics, Zhengzhou University, Zhengzhou 450001, China}

\author[0009-0000-3578-9134]{Bingqing Sun}
\affiliation{Department of Astronomy, University of Massachusetts Amherst, 710 North Pleasant Street, Amherst, MA 01003, USA}

\author[0000-0001-8382-6323]{Shi Shao}
\affiliation{Key Laboratory for Computational Astrophysics, National Astronomical Observatories, Chinese Academy of Sciences, 20A Datun Road, Chaoyang District, Beijing 100101, China}
\affiliation{School of Astronomy and Space Science, University of Chinese Academy of Sciences, Beijing 100049, China}

\correspondingauthor{Shihong Liao} \email{shliao@nao.cas.cn}

\begin{abstract}
Traditional cosmological hydrodynamical simulations usually assume equal-numbered but unequal-mass dark matter and baryonic particles, which can lead to spurious collisional heating due to energy equipartition. To avoid such a numerical heating effect, a simulation setup with equal-mass dark matter and baryonic particles, which corresponds to a particle number ratio of $N_{\rm DM}:N_{\rm gas} = \Omega_{\rm cdm} / \Omega_{\rm b}$, is preferred. However, previous studies have typically used grid-based particle loads to prepare such initial conditions, which can only reach specific values for $N_{\rm DM}:N_{\rm gas}$ due to symmetry requirements. In this study, we propose a method based on the glass approach that can generate two-component particle loads with more general $N_{\rm DM}:N_{\rm gas}$ ratios. The method simultaneously relaxes two Poisson particle distributions by introducing an additional repulsive force between particles of the same component. We show that the final particle load closely follows the expected minimal power spectrum, $P(k) \propto k^{4}$, exhibits good homogeneity and isotropy properties, and remains sufficiently stable under gravitational interactions. Both the dark matter and gas components individually also exhibit uniform and isotropic distributions. We apply our method to two-component cosmological simulations and demonstrate that an equal-mass particle setup effectively mitigates the spurious collisional heating that arises in unequal-mass simulations. Our method can be extended to generate multi-component uniform and isotropic distributions. Our code based on \textsc{gadget-2} is available at \href{https://github.com/liaoshong/gadget-2glass}{https://github.com/liaoshong/gadget-2glass}.
\end{abstract}


\section{Introduction} \label{sec:intro}

Preparing a uniform (and preferably isotropic) \textit{particle load} (or \textit{pre-initial condition}) is the first crucial step in setting up the initial condition for a cosmological simulation \citep[see, e.g.,][for a review]{Angulo2022}. The particle load represents the uniform and isotropic state of the Universe at $a = 0$ and sets the initial state of the particle distribution in simulations before perturbations, specified by the input matter power spectrum, are imposed. Ensuring that this initial distribution is both uniform and isotropic helps minimize numerical artifacts. This ensures that any structure formation observed in the simulation originates from the input physical perturbations and is driven by gravitational dynamics, rather than by artificial noises introduced during the setup. As a result, the preparation of particle loads affects the accuracy and reliability of cosmological simulations, especially when investigating the formation of cosmic structures across different scales and involving multiple components.

For simulations containing a single component (e.g., pure dark matter N-body simulations), the known particle loads include grid \citep[e.g.,][]{Efstathiou1985}, glass \citep{White1996}, quaquaversal tiling \citep[or Q-set,][]{Hansen2007}, and Capacity Constrained Voronoi Tessellation \citep[CCVT,][]{Liao2018}. Many analytical and numerical work have investigated the impact of particle loads on the growth of dark matter structures and the convergence among different loads \citep[see e.g.,][]{Baugh1995,Gotz2002,Gotz2003,Smith2003,Joyce2005,Joyce2009,Marcos2006,Joyce2007,Wang2007,Schmittfull2013,LHuillier2014,Garrison2016,Liao2018,Masaki2021,Zhang2021,Bagla2024,SylosLabini2025,Yu2025}.

For simulations with two components (e.g., dark matter and gas), the classical approach assumes that one component perfectly follows the other (i.e., both dark matter and gas follow the transfer function of total matter). Under this assumption, a single set of particles representing the total matter (dark matter plus gas) distribution is first generated using the same approach as in the single-component case. After the perturbations are added, this set of particles is then split into two sets (one for dark matter and the other for gas), with particles displaced in opposite directions. The total displacement is equal to half of the mean particle separation in each dimension, ensuring the center of mass remains unchanged. In this setup, dark matter and gas each have the same number of particles but different particle masses, specified according to the cosmological parameters $\Omega_{\rm m}$ and $\Omega_{\rm b}$. Here, $\Omega_{\rm m}$ and $\Omega_{\rm b}$ denote the density fractions of total matter and baryonic matter, respectively, relative to the critical density. For instance, in the Planck cosmology \citep{Planck_2020}, the dark matter particle mass is roughly five times that of the gas particle mass. This method is implemented in the \textsc{Ngenic} code \citep{Springel2005Nature,Angulo2012}, which has been widely used for setting up cosmological hydrodynamical simulations. Interested readers are referred to \citet{Liao2017} for a detailed description and an application to study the segregation effect of dark matter and baryons.

However, in recent years, it has been realized that the use of equal-number but unequal-mass particle sets can lead to spurious collisional heating \citep{Ludlow2019}. Specifically, more massive dark matter particles transfer their kinetic energy to lighter baryonic particles, as a result of the equipartition of energy in a gravitational system. This `heating' effect is numerical in nature, arising from the unequal mass elements used in simulations. Spurious heating can affect galaxy properties, including sizes, morphologies, dark matter and stellar kinematics, and the thickness of galactic discs \citep[see e.g.,][]{Ludlow2019,Ludlow2021,Ludlow2023,Wilkinson2023,Zeng2024}.

A solution to avoid spurious heating is to use identical particle masses for dark matter and baryonic particles. Equivalently, different numbers of particles are adopted in the initial condition, with the ratio given by $N_{\rm DM}/N_{\rm gas} = (\Omega_{\rm m} - \Omega_{\rm b}) / \Omega_{\rm b}$. Under Planck cosmology \citep{Planck_2020}, this ratio is approximately $5.35:1$. This requires preparing separate initial particle loads for dark matter and gas. \citet{Richings2021} propose a method of creating a $N_{\rm DM}:N_{\rm gas} = 7:1$ load, which closely approximates the Planck cosmology ratio. This approach tessellates a cubic template where one gas particle is placed at the center, and 26 dark matter particles are symmetrically distributed across the six faces, twelve edges, and eight vertices. Since each face is shared by two templates, each edge by four, and each vertex by eight, the effective dark matter count per template is $6/2 + 12/4 + 8/8 = 7$. This approach has been used in the studies of \citet{Ludlow2019,Ludlow2023} and the MAGPIE simulations (Shao et al. in prep.). A similar method is adopted in the recent COLIBRE simulations \citep{Schaye2025}, where 14 dark matter particles are placed at the six face centers and eight vertices of the template, yielding an effective dark matter particle number per template of $6/2 + 8/8 = 4$, and thus a number ratio $N_{\rm DM} : N_{\rm gas} = 4:1$. In \citet{Bird2020}, the authors test with an initial condition setup using two offset grids with $N_{\rm DM} = 256^3$ and $N_{\rm gas} = 150^3$, achieving a ratio of $N_{\rm DM}:N_{\rm gas} = 4.97:1$. All these methods and their variants produce grid-based particle loads and can only achieve specific values for $N_{\rm DM}/N_{\rm gas}$ due to symmetry requirements.

In this work, we aim to address the following question: {\it Can a two-component particle load be prepared to achieve a more general $N_{\rm DM}:N_{\rm gas}$ ratio?} If such a solution exists, we would be able to ensure identical particle masses for dark matter and gas under different cosmologies. In addition to avoiding spurious heating, the equal-mass approach increases the mass resolution of dark matter and thus improves the ability to resolve dark matter subhalos in cosmological simulations compared to the classical unequal-mass approach \citep[see][]{Richings2021}. Moreover, in simulations that include elastic interactions between different components, having equal-mass particles or adopting more general $N_{\rm DM}:N_{\rm gas}$ ratios is also crucial (\citealt{Fischer2025}; Zhang et al., in prep.).

This paper is structured as follows. In Section~\ref{sec:met}, we describe our method for generating two-component particle loads. The properties of our particle loads are quantified in Section~\ref{sec:pro}. The application of our particle loads in cosmological simulations is explored in Section~\ref{sec:app}. Finally, we summarize in Section~\ref{sec:sum}. Some detailed tests are presented in appendices.

\section{Methods} \label{sec:met}

\subsection{Background and considerations}

Compared to the symmetric grid-based particle load, glass is a method for constructing a uniform and isotropic distribution with any desired number of particles. Therefore, we consider glass-based particle loads in this study.

A single-component glass load is generated by evolving a Poisson particle distribution\footnote{This refers to a distribution where particle positions are drawn independently from a uniform probability distribution over the volume.} under anti-gravity \citep{White1996}. Specifically, at each step, the gravitational force acting on each of the $N_{\rm p}$ particles is computed as follows:
\begin{equation}\label{eq:single_force}
    \vect{F}_i = \sum_{j = 1, j \neq i}^{N_{\rm p}} G m_i m_j \frac{\vect{r}_j - \vect{r}_i}{|\vect{r}_{j} - \vect{r}_i|^3}, 
\end{equation}
where $G$ is the gravitational constant, $m_i$ and $\vect{r}_i$ are the mass and position of particle $i$. The particles are then evolved under the force of $-\vect{F}_i$ until the entire system reaches a (quasi-)equilibrium state. 

For two-component glass-based particle loads, \citet{Yoshida2003} conducted a detailed study on the case of $N_{\rm DM}:N_{\rm gas} = 1:1$ when they explored using different transfer functions for baryons and dark matter in initial conditions.\footnote{The rationale behind \citet{Yoshida2003} is that baryons and dark matter decouple from photons at different epochs, leading to distinct evolutions in their power spectra at high redshifts. To precisely capture these differences in two-fluid cosmological simulations, careful improvements on the classical approach mentioned in Section~\ref{sec:intro}, which assumes that two fluids perfectly follow each other in initial conditions, are necessary. For further discussions on this topic, see e.g., \citet{Angulo2013,Valkenburg2017,Bird2020,Hahn2021,Liu2023}.} They suggest independently generating two glass distributions, combining them, and then evolving the combined distribution under anti-gravity for a certain number of steps to avoid close particle juxtapositions.

Note that the last step in the \citet{Yoshida2003} method is necessary because, when two independently prepared glass loads are combined, particles may end up positioned too close to one another. However, the subsequent anti-gravity evolution may then cause deviations from the original uniformity of each component. To improve this, one possible approach is to relax the two components simultaneously rather than independently.

In the following subsection, we generalize the ideas from \citet{White1996} and \citet{Yoshida2003} by simultaneously relaxing two Poisson particle distributions to generate two-component glass-based particle loads. This method allows for arbitrary $N_{\rm DM}{:}N_{\rm gas}$ ratio while maintaining good uniformity, isotropy, and force-free properties.

\subsection{Two-component glass-based particle loads}\label{subsec:met_details}

We first generate two Poisson particle sets, one consisting of $N_{\rm gas}$ particles and the other consisting of $N_{\rm DM}$ particles. Each particle in the gas\footnote{Although we refer to this particle set as {\it gas} for convenience, hydrodynamics is not considered when preparing the particle loads. Both gas and dark matter particles are treated as collisionless when generating particle loads.} component is then evolved under the force of
\begin{eqnarray}\label{eq:gas_force}
    -\vect{F}_{i{\rm(gas)}} &=& - \sum_{j = 1, j \neq i}^{N_{\rm tot}} G m_i m_j \frac{\vect{r}_j - \vect{r}_i}{|\vect{r}_{j} - \vect{r}_i|^3} \nonumber \\
    &~& - C_{\rm gas} \sum_{j = 1, j \neq i}^{N_{\rm gas}} G m_i m_j \frac{\vect{r}_j - \vect{r}_i}{|\vect{r}_{j} - \vect{r}_i|^3}, 
\end{eqnarray}
where the first term on the right-hand side sums over all particles ($N_{\rm tot} = N_{\rm gas} + N_{\rm DM}$), while the second term sums only over gas particles. Note that $m_i = m_j$ since we are considering equal-mass particles. In the second term, the factor $C_{\rm gas} = (N_{\rm tot}/N_{\rm gas})^{2/3}$ rescales the gas particle system, ensuring that the forces from the entire particle system and the gas particle system alone are more comparable (see Appendix~\ref{ap:force_term} for further discussion of the choice of this factor). Thus, the first term accounts for the anti-gravity from all particles, while the second term computes the scaled anti-gravity from the gas particles only.

Similarly, each particle in the dark matter component is evolved under the force of 
\begin{eqnarray}\label{eq:dm_force}
    -\vect{F}_{i{\rm(DM)}} &=& - \sum_{j = 1, j \neq i}^{N_{\rm tot}} G m_i m_j \frac{\vect{r}_j - \vect{r}_i}{|\vect{r}_{j} - \vect{r}_i|^3} \nonumber \\
    &~& - C_{\rm DM} \sum_{j = 1, j \neq i}^{N_{\rm DM}} G m_i m_j \frac{\vect{r}_j - \vect{r}_i}{|\vect{r}_{j} - \vect{r}_i|^3}, 
\end{eqnarray}
where $C_{\rm DM} = (N_{\rm tot} / N_{\rm DM})^{2/3}$ (see Appendix~\ref{ap:force_term} for more details), and the second term on the right-hand side is the rescaled anti-gravity from the dark matter particles only.

Compared to the single-component glass case (Equation~\ref{eq:single_force}), the two-component system introduces an additional repulsive force within each individual component. Intuitively, this repulsive force causes particles of the same component to move away from one another. When the entire particle system reaches a (quasi-)equilibrium state, all particles are distributed in a way that maximizes separation. At the same time, particles within the same component avoid each other as much as possible. Since gravity has no preferred direction, both the total configuration and the individual component configurations achieve uniformity and isotropy.

\begin{figure*} 
\centering\includegraphics[width=0.9\textwidth]{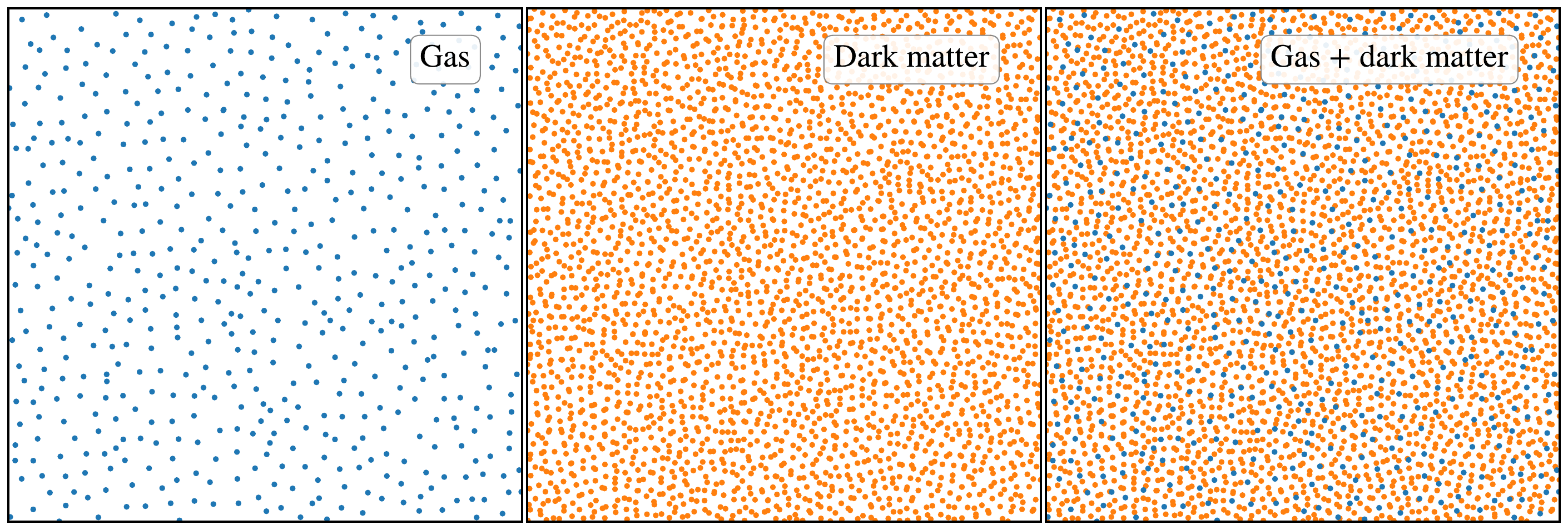}
\caption{Visualization of a two-component glass-based particle load consisting of $24^3$ gas particles and $73945$ $(\approx 42^3)$ dark matter particles (i.e., $N_{\rm DM}:N_{\rm gas} = 5.35:1$). From left to right, the panels show the gas (blue), dark matter (orange), and total particles within a slice of thickness $L_{\rm box}/24$, projected onto the $xy$-plane. The particle distributions visually demonstrate that both individual components and the entire particle set exhibit glass-like characteristics, i.e., overall uniformity and isotropy.}
\label{fig:visua}
\end{figure*}

To achieve better force balance across the entire particle distribution, we switch off the additional force from the same component (i.e., the second term on the right-hand side of Equations~\ref{eq:gas_force} and \ref{eq:dm_force}), and evolve the system as a single glass for some steps (see Appendix~\ref{ap:steps} for more details). Specifically, all particle loads discussed in Sections~\ref{sec:pro} and \ref{sec:app} were generated by evolving in total $2^{13} = 8192$ steps, with the last $20$ steps switching off the additional force. Since the two-component glass-based distribution is already fairly uniform, this additional evolution only perturbs the system slightly, preserving the uniformity and isotropy of each component.

Similar to the single-component glass load, this method can easily incorporate the periodic boundary condition. Therefore, we only need to prepare a small particle load, and then use the tiling method to obtain a large particle load.

Compared to the approaches in \citet{Yoshida2003} and \citet{Bird2020}, which combine two independent particle loads and potentially result in close juxtapositions between dark matter and gas particles, our method evolves two particle sets simultaneously from their initial random distributions, naturally avoiding such overlaps. Consequently, the subsequent anti-gravity evolution of the entire particle system has less impact on each individual set, resulting in a more uniform distribution within each (see Appendix~\ref{ap:compare} for quantitative comparisons).

We have implemented this method into the \textsc{gadget-2} code \citep{Springel2005}, and it is publicly available at \href{https://github.com/liaoshong/gadget-2glass}{https://github.com/liaoshong/gadget-2glass}.

\section{Properties of particle loads} \label{sec:pro}

\subsection{Visualization}\label{subsec:visua}

In Figure~\ref{fig:visua}, we show a two-component glass-based particle load generated using the method outlined in Section~\ref{subsec:met_details}. For better visualization, we present a particle load with fewer particles than commonly used. It consists of $24^{3}$ gas particles and $73945 ~ (\approx 42^3)$ dark matter particles, achieving a particle number ratio following the cosmological parameters from the Planck 2018 results \citep{Planck_2020}, i.e.,
\begin{equation}
    \frac{N_{\rm DM}}{N_{\rm gas}} = \frac{\Omega_{\rm cdm}}{\Omega_{\rm b}} \approx \frac{\Omega_{\rm m} - \Omega_{\rm b}}{\Omega_{\rm b}} \approx 5.35, \footnote{The first approximately equals sign comes from the fact that in the Planck 2018 results, $\Omega_{\rm m}$ includes the contribution from neutrinos with a mass of $0.06~{\rm eV}/c^{2}$. See \citet{Planck_2020} for details.} 
\end{equation}
where $\Omega_{\rm cdm}$ denotes the density fraction of cold dark matter with respect to the critical density, and we adopt $\Omega_{\rm m} = 0.3111$, $\Omega_{\rm b} = 0.04897$, and $h = 0.6766$.

A visual inspection of the individual components and the entire particle set suggests that they exhibit glass-like characteristics, i.e., overall uniformity and the absence of preferred directions. Some particles may appear close to one another, but this is due to projection effects, as we are visualizing a slice of thickness $L_{\rm box}/24$, where $L_{\rm box}$ denotes the side length of the periodic cubic box. In Section~\ref{subsec:uni_iso}, we will quantitatively demonstrate that our two-component particle loads possess relatively good uniformity and isotropy properties.

\subsection{Power spectra}\label{subsec:pk}

\begin{figure} 
\centering\includegraphics[width=0.48\textwidth]{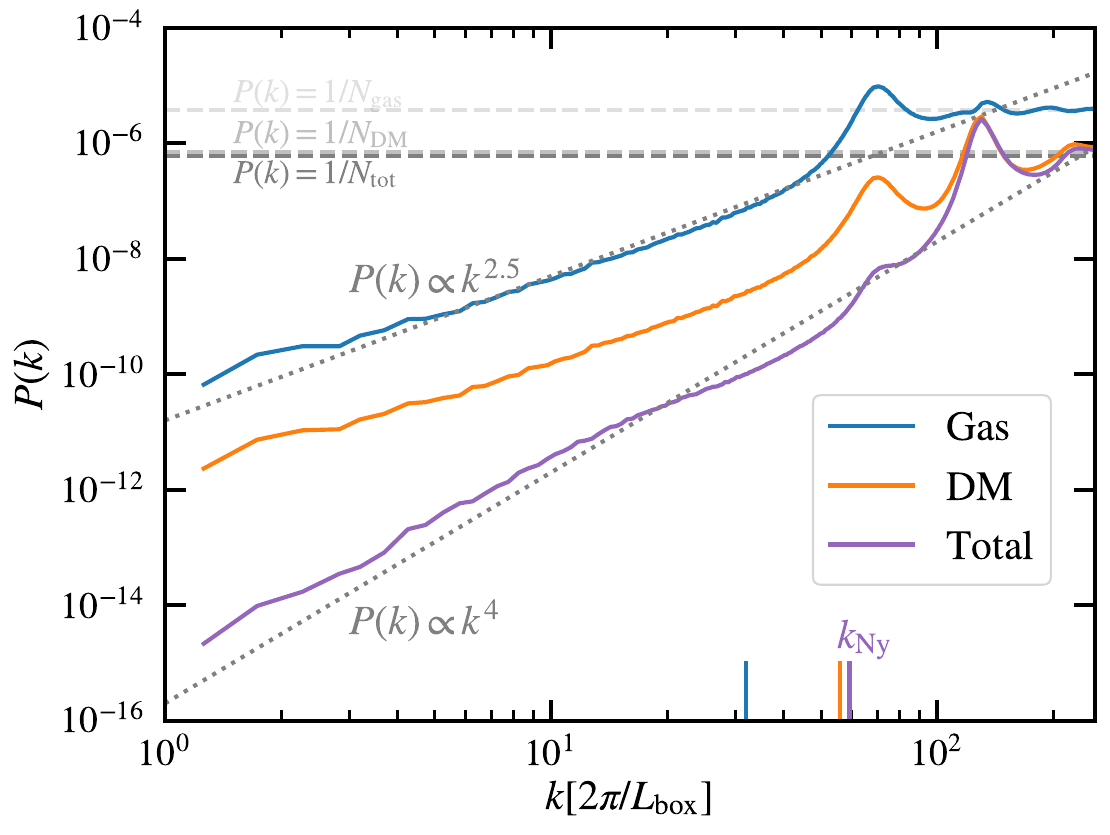}
\caption{Power spectra of a two-component glass-based particle loads with $N_{\rm gas} = 64^3$ and $N_{\rm DM} = 1402203 \approx 112^3$ (i.e., $N_{\rm DM}:N_{\rm gas} = 5.35:1$). The blue, orange, and purple curves show the power spectra of the gas particles, and the dark matter particles, and the entire set respectively. The horizontal dashed lines show the Poisson noise power spectra. The dotted lines represent the power-law power spectra; the upper dotted line is for $P(k) \propto k^{2.5}$, and the lower one is for the minimal power spectrum, $P(k) \propto k^4$. The vertical line segments at the bottom mark the particle Nyquist frequencies ($k_{\rm Ny}$) for different particle sets. Overall, all power spectra approximately follow the minimal power spectrum at scales below $k_{\rm Ny}$ and gradually become dominated by Poisson noise at $k \gtrsim k_{\rm Ny}$, demonstrating glass-like properties.}
\label{fig:pk}
\end{figure}

In this and the following subsections, we consider a particle load with a larger number of particles, which is more commonly used. Specifically, we use $N_{\rm gas} = 64^3$ gas particles and $N_{\rm DM} = 1402203 ~ (\approx 112^3)$, again following the baryon-to-dark matter ratio given by the Planck 2018 results.

The power spectra of each component and the total particle set are plotted in Figure~\ref{fig:pk}. Below the particle Nyquist frequency $k_{\rm Ny}$, the total power spectrum closely follows the expected minimal power spectrum, $P(k) \propto k^{4}$ \citep{Zeldovich1965,Peebles1980}, which is a characteristic of glass distributions. Above $k_{\rm Ny}$, the total power spectrum gradually becomes dominated by Poisson noise. Similar behavior is observed in the power spectra of gas and dark matter components, although at large scales (i.e., $k \la k_{\rm Ny}$), the slope of $P(k)$ is somewhat shallower (i.e., $\propto k^{2.5}$). The dark matter power spectrum exhibits a lower peak between $k_{\rm Ny}$ and the first peak where it reaches the Poisson noise level. This feature is likely influenced by the gas particle distribution, as the scale of this peak coincides with that of the first Poisson noise peak in the gas power spectrum. Intuitively, the dark matter particle distribution is analogous to removing a uniform and isotropic subset of particles from a glass distribution, and these `holes' imprint a characteristic correlation at the scale corresponding to their mean separation, which manifests as the lower peak observed here. 

Overall, all power spectra, including those of individual components and the entire particle set, exhibit glass-like properties, indicating that the particle distributions closely resemble glass configurations.

\subsection{Homogeneity and isotropy}\label{subsec:uni_iso}

\begin{figure} 
\centering\includegraphics[width=0.48\textwidth]{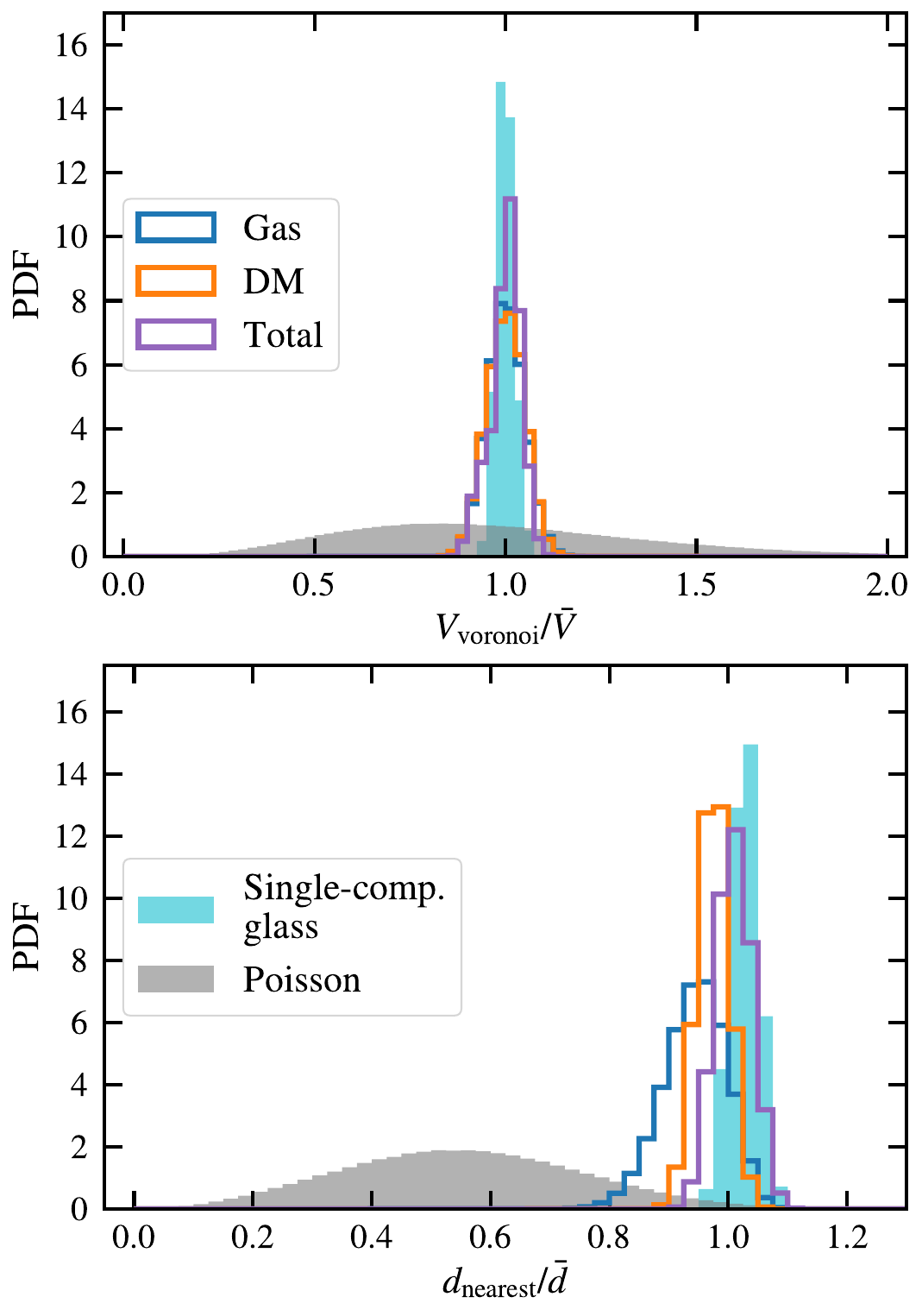}
\caption{Homogeneity properties. The upper panel shows the PDFs of the Voronoi cell volume associated with each particle. As before, gas particles, dark matter particles, and the entire particle set are represented in blue, orange, and purple, respectively. The Voronoi cell volumes are normalized to the mean particle volume for each particle set, i.e., $\bar{V} = L_{\rm box}^3 / N_{i}$ with $i = $ gas, DM, or total. For comparison, the results from a single-component glass load and a Poisson particle distribution, both generated with the same total number of particles $N_{\rm DM} + N_{\rm gas}$, are shown in cyan and gray, respectively. The lower panel displays similar PDFs but for the distance to the nearest neighbor. Here, the distances are normalized to the mean inter-particle separation, i.e., $\bar{d} = L_{\rm box} / N_{i}^{1/3}$. From this figure, particles in all three sets occupy space evenly, with their nearest-neighbor distances close to the mean inter-particle separation, indicating that the distributions are relatively uniform.}
\label{fig:uniformity}
\end{figure}

\begin{figure} 
\centering\includegraphics[width=0.48\textwidth]{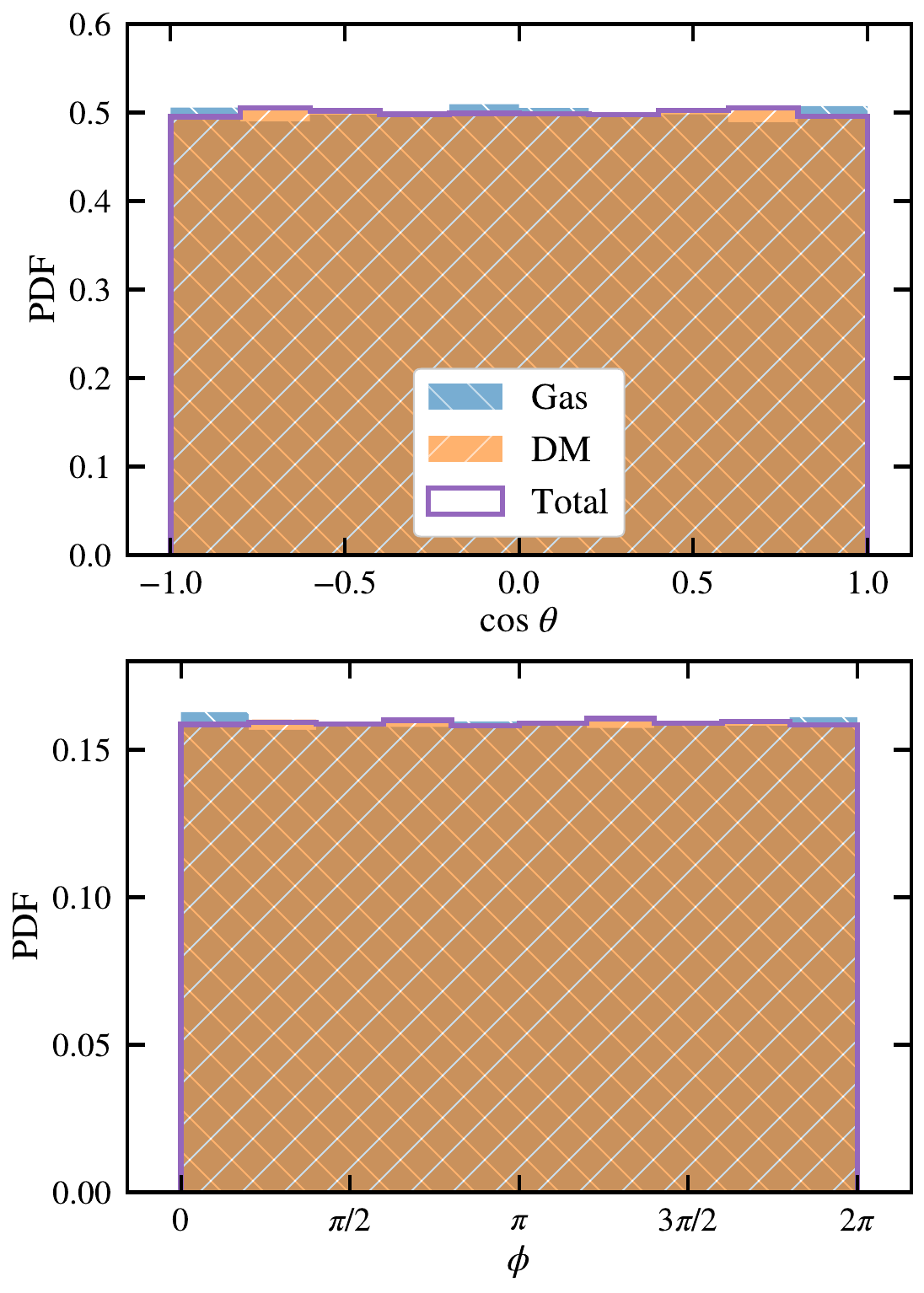}
\caption{Isotropy properties. To quantify the isotropy of particle loads, we determine the vector pointing from each particle to its nearest neighbor and compute the polar angle $\theta$ and the azimuthal angle $\phi$ (with respect to the Cartesian coordinate axes of the periodic box). The upper panel shows the PDFs of $\cos \theta$ for gas particles (blue), dark matter particles (orange), and the entire particle set (purple). Similarly, the lower panel displays that PDFs of the azimuthal angle $\phi$. Both the distributions of $\cos \theta$ and $\phi$ are uniform, indicating that there is no preferred direction in the particle distributions.}
\label{fig:isotropy}
\end{figure}

In this subsection, we further quantify the homogeneity and isotropy properties of our particle loads. To measure homogeneity, we consider two quantities: the Voronoi cell volume, $V_{\rm voronoi}$, and the distance to the nearest neighbor, $d_{\rm nearest}$.

{\it Voronoi cell volume.} For each particle set (i.e., gas, dark matter, and total), we perform a Voronoi tessellation accounting for periodic boundary conditions and compute $V_{\rm voronoi}$ for each particle. The probability distribution functions (PDFs) of the gas, dark matter, and total particle set are plotted in the top panel of Figure~\ref{fig:uniformity}. For easy comparison, we normalize $V_{\rm voronoi}$ to the mean particle volume of each particle set, $\bar{V}_{i} = L_{\rm box}^3 / N_{i}$, where $i = $ gas, DM, or total. We also include the PDFs from a single-component glass load (cyan) and a Poisson distribution (gray), both generated with $N_{\rm tot}$ particles. We can see that, similar to the single-component glass distribution, both the individual components and the total particle set in our particle load exhibit a relatively narrow distribution peaking at $V_{\rm voronoi}/\bar{V} = 1$. The PDFs of the dark matter and gas components are slightly less narrow compared to that of the total particle set. Both the individual and total particle sets stand in contrast to the Poisson distribution, which has a much broader PDF due to the clustering of some particles in a Poisson process \citep[for further discussions of Poisson Voronoi diagrams, see][]{Okabe2000}. This indicates that all three particle sets in our load occupy space evenly and thus their distributions are fairly homogeneous.

{\it Distance to the nearest neighbor.} For each particle in each set, we identify its nearest neighbor from the same set using a k-d tree \citep{Bentley1975}, incorporating periodic boundary conditions, and compute their separation $d_{\rm nearest}$. We normalize $d_{\rm nearest}$ to the mean inter-particle separation of each set, $\bar{d}_{i} = L_{\rm box} / N_{i}^{1/3}$ where $i = $ gas, DM, and total. The PDFs of $d_{\rm nearest} / \bar{d}$ for different particle sets are displayed in the bottom panel of Figure~\ref{fig:uniformity}. The PDF of the total particle set closely resembles that of the single-component glass, with a narrow distribution around $1$. The PDF of the dark matter component peaks slightly below $1$, while the PDF of the gas component is somewhat broader. Again, this contrasts sharply with the Poisson particle distribution, which is significantly broader and has a mean value of ${\sim}0.55$ \citep[see][for an analytical derivation]{Chandrasekhar1943}. Overall, all particle sets tend to have their nearest-neighbor distances close to the mean inter-particle separation, indicating that their distributions are relatively uniform.

To quantify isotropy, for each particle set, we consider the unit vector pointing from each particle to its nearest neighbor and compute the polar angle $\theta$ and the azimuthal angle $\phi$ with respect to the Cartesian coordinate axes of the periodic box. If there is no preferred direction in a particle set, the distributions of $\cos \theta$ and $\phi$ should be uniform. Figure~\ref{fig:isotropy} presents the PDFs of $\cos \theta$ and $\phi$ for both the individual components and the entire particle set. All PDFs are nearly flat, confirming that both the individual particle sets and the total particle set exhibit excellent isotropy.

\subsection{Force balance}\label{subsec:for_bal}

\begin{figure} 
\centering\includegraphics[width=0.48\textwidth]{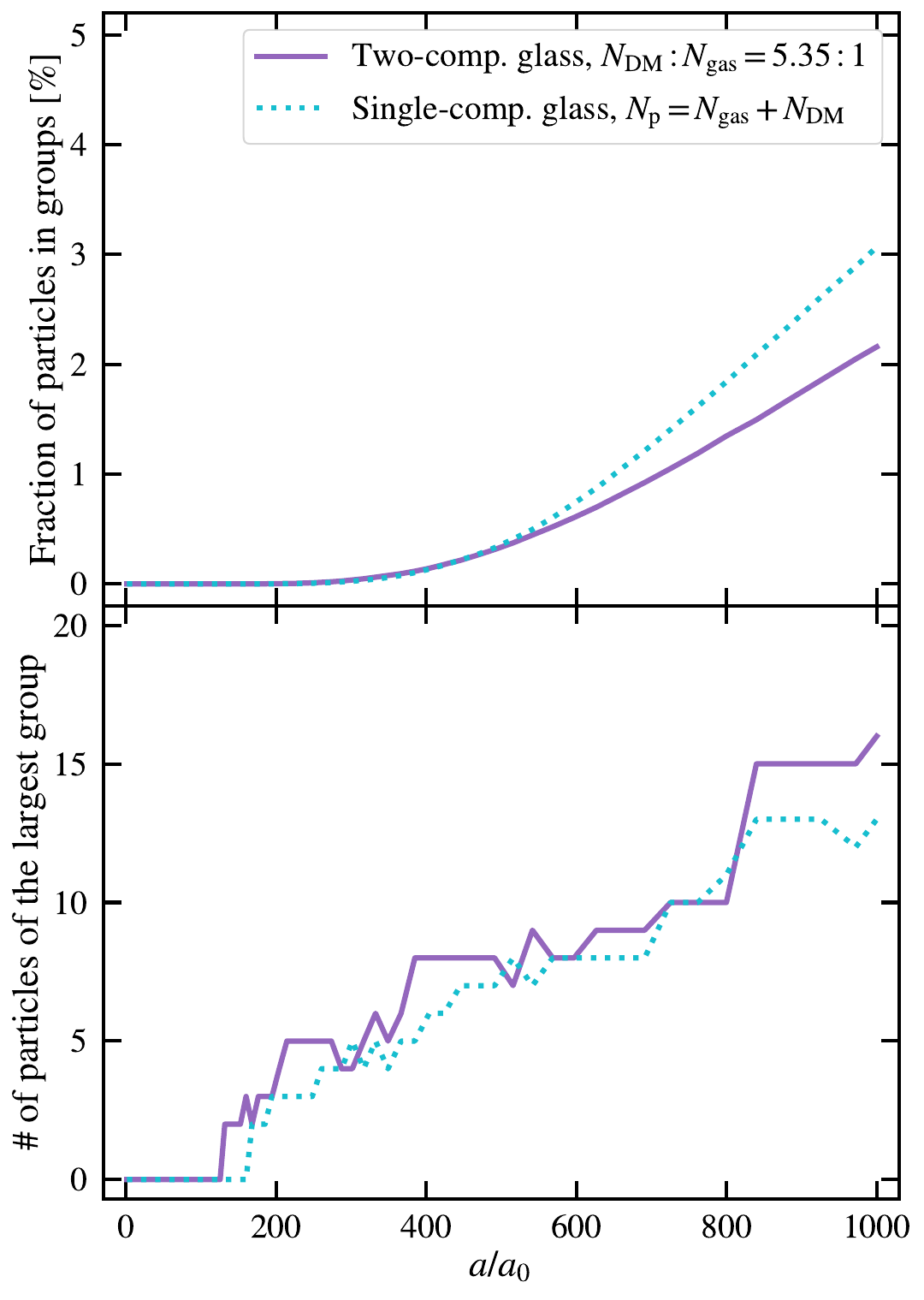}
\caption{Property of force balance. The entire particle load is evolved under gravitational interactions from $a_0 = 0.001$ to $a = 1$ within the SCDM cosmology. The top panel shows that fraction of particles in the identified FOF groups as a function of the expansion factor ($a / a_0$), whereas the bottom panel plots the number of particles for the largest FOF group in different snapshots. The purple curves represent the two-component particle loads with $N_{\rm DM} : N_{\rm gas} = 5.35:1$. For comparison, the traditional single-component glass load with identical total number of particles is plotted with cyan dotted lines. The two-component particle load exhibits behavior similar to that of the traditional single-component glass load and is fairly stable to prevent artificial structures from forming due to noise.}
\label{fig:force_balance}
\end{figure}

A high-quality particle load should also be force-free, i.e., each particle in the load experiences zero net gravitational force from other particles. This ensures that the intrinsic distribution itself does not develop structures under gravity and the structures that form in a simulation originate from the physical perturbations imposed onto the load. 

To examine the force-balance property of our particle load, we evolve the entire load in the standard cold dark matter (SCDM, with $\Omega_{\rm m} = 1$) model using the \textsc{gadget-2} code \citep{Springel2005}. Only gravitational interactions, computed using the TreePM method \citep{Xu1995,Springel2005}, are considered in the simulation. The simulation begins at $a_0 = 0.001$ and runs until $a = 1$, corresponding to an expansion by a factor of $1000$. A total of $143$ snapshots are saved throughout the evolution. In each snapshot, we utilize the friends-of-friends \citep[FOF,][]{Davis1985} algorithm with a linking length parameter of $b = 0.2$ to identify `structures' containing at least 2 particles.

The fraction of particles in all identified FOF groups as a function of time is shown by the purple solid line in the top panel of Figure~\ref{fig:force_balance}, while the number of particles of the largest FOF group over time is presented in the bottom panel. For comparison, we have also evolved a single-component glass load with $N_{\rm tot}$ particles and performed a parallel analysis, with the results displayed using cyan dotted lines. As seen in the figure, our two-component particle load exhibits behavior very similar to the traditional single-component glass load. FOF groups only start to form after the universe has expanded by a factor of ${\sim}150$, and even by the end of the simulations, only a few per cent of particles are found in these groups and the largest group only contains ${\sim}15$ particles. In practical simulations, the universe typically expands by a factor of ${\sim}100$ (i.e., from $z {\sim} 100$ to $z = 0$). Therefore, our particle loads should be sufficiently stable to prevent artificial structures from forming due to noise.

\section{Applications} \label{sec:app}

\begin{figure*} 
\centering\includegraphics[width=\textwidth]{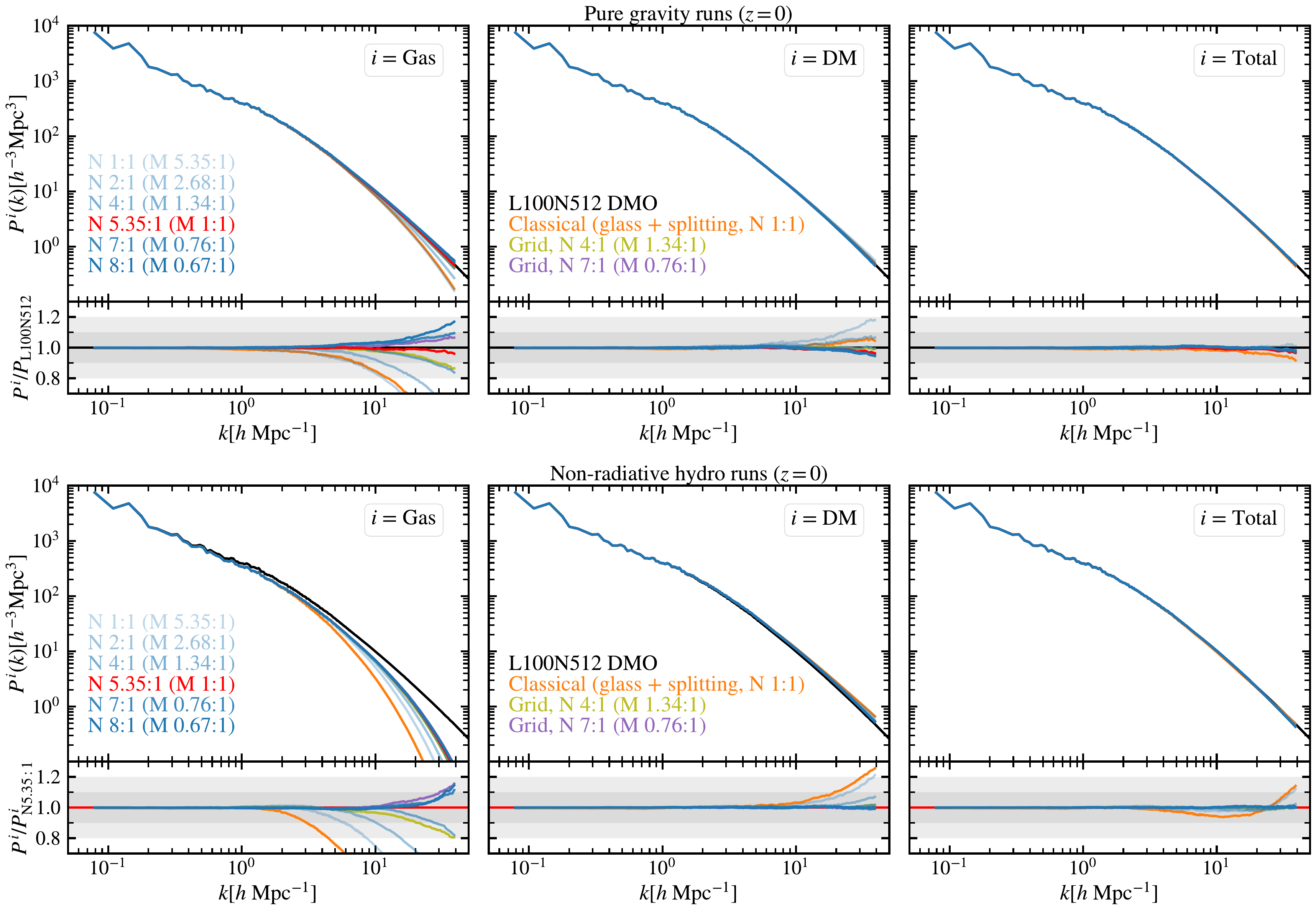}
\caption{Impact of particle mass ratios on matter power spectra at $z=0$. {\it Top}: Results from pure-gravity simulations. From left to right, the panels display the power spectra for gas, dark matter, and their total. In all panels, the black line represents a standard dark-matter-only run (L100N512 DMO) for reference. The orange line shows a run using the classical particle-splitting initialization. The olive and purple lines plot runs using the $4:1$ \citep{Schaye2025} and $7:1$ \citep{Richings2021} grid particle loads, respectively. All other colored lines represent simulations using the two-component particle loading method introduced in this work, with the legend indicating the number (N $N_{\rm DM}:N_{\rm gas}$) and mass (M  $m_{\rm DM}:m_{\rm gas}$) ratios for each. In particular, the run with equal-mass dark matter and gas particles is shown in red. The subpanels show the ratio of each power spectrum to the dark-matter-only run. {\it Bottom}: Similar to top panels, but showing results from non-radiative hydrodynamical simulations. Note that unlike the top panels, here we plot the ratio of each power spectrum to that of the equal-mass run (red line). These tests reveal a systematic effect in unequal-mass simulations: the power spectrum of the lighter particle component is suppressed at small scales, while that of the heavier component is enhanced. This artifact arises from spurious collisional heating as kinetic energy is transferred from massive to lighter particles.}
\label{fig:application_pk}
\end{figure*}

\subsection{Simulation setups}

In this section, we apply our new particle loading method to cosmological simulations and compare the results with those from traditional methods. We run a set of $\Lambda$CDM simulations using the Planck 2018 cosmological parameters\footnote{Apart from the parameters outlined in Section~\ref{subsec:visua}, others are $\Omega_\Lambda = 0.6889$, $\sigma_8 = 0.8102$, and $n_{\rm s} = 0.9665$.} \citep{Planck_2020}, with the following initial particle loads:

(i) Classical approach. A single glass particle load is used to generate the total matter distribution. The particles are then split into two sets by translating them in opposite directions, yielding a particle number ratio of $N_{\rm DM}:N_{\rm gas} = 1:1$ and a mass ratio of $m_{\rm DM} : m_{\rm gas} = 5.35:1$.

(ii) $4:1$ grid particle load \citep{Schaye2025}. This results in ratios of $N_{\rm DM}:N_{\rm gas} = 4:1$ and $m_{\rm DM} : m_{\rm gas} = 1.34:1$.

(iii) $7:1$ grid particle load \citep{Richings2021}. This results in ratios of $N_{\rm DM}:N_{\rm gas} = 7:1$ and $m_{\rm DM} : m_{\rm gas} = 0.76:1$.

(iv) Our method. We generate particle loads with various number ratios, $N_{\rm DM}:N_{\rm gas} = 1:1$, $2:1$, $4:1$, $5.35:1$, $7:1$, and $8:1$. These correspond to mass ratios of $m_{\rm DM} : m_{\rm gas} = 5.35:1$, $2.68:1$, $1.34:1$, $1:1$, $0.76:1$, and $0.67:1$, respectively. The $1:1$, $4:1$, and $7:1$ runs are designed with the same particle numbers as methods (i) -- (iii) for direct comparison.

The initial redshift is $z_{\rm IC} = 127$ for all runs. We generate the initial condition for the classical approach using the \textsc{Ngenic} code \citep{Springel2005Nature,Angulo2012}. For all other runs, we use a modified version of this code in which the gas and dark matter displacement fields are generated from the same power spectrum (i.e., the total matter power spectrum) and random phases. Each simulation contains $256^3$ gas particles, with the number of dark matter particles set by the ratios outlined above. The simulation box size is $L_{\rm box} = 100~h^{-1}{\rm Mpc}$ and the gravitational softening length is set to $\epsilon = 3.9~h^{-1}{\rm kpc}$ for both particle types, which corresponds to 1/100 of the mean inter-gas particle separation. All simulations in this section are evolved using the \textsc{gadget-2} code \citep{Springel2005}.

For comparison, we also perform a higher-resolution, dark-matter-only (DMO) simulation (L100N512), containing $512^3$ particles. In this run, the gravitational softening length is set to $1.95~h^{-1}{\rm kpc}$, while other simulation parameters are kept identical to the main runs.

\subsection{Pure gravity runs}\label{subsec:app_pure_grav}

To investigate how an equal-mass particle setup mitigates spurious collisional heating in two-component cosmological simulations, we first perform a set of pure gravity runs. The resulting power spectra for the gas, dark matter, and total components at $z = 0$ are shown in the top row of Figure~\ref{fig:application_pk}. Since both components start with the same initial power spectrum and are evolved under gravity alone, they are expected to produce identical power spectra at $z = 0$ in the absence of other numerical artifacts. Furthermore, on scales where the simulations are converged, these spectra should match the result from our higher-resolution L100N512 DMO run.\footnote{We determined this convergence scale by running an additional L100N256 DMO simulation, which has the same particle number ($256^3$) as the gas component in our main runs. Its $z=0$ power spectrum converges to that of the L100N512 run at a level of $\leq 10\%$ for $k \lesssim 40~h~{\rm Mpc}^{-1}$. We therefore plot the power spectra of our main runs in this $k$-range in Figure~\ref{fig:application_pk}.}

As shown in the top panels of Figure~\ref{fig:application_pk}, only the equal-mass run (red line) meets this expectation. In this case, both the gas and dark matter power spectra closely follow the result of the L100N512 run, with their ratio being very close to unity. For the unequal-mass runs, the component with lighter particles has its small-scale power suppressed, while the component with more massive particles has its power enhanced. It is worth noting, however, that the total matter power spectrum is less sensitive to these effects, with all runs agreeing with each other to within 10\%.

This behavior arises from spurious collisional heating, where kinetic energy is transferred from massive to lighter particles, suppressing small-scale structures in the lighter component while enhancing those in the massive one. This artifact is more pronounced in the gas component, as its smaller number of particles means the energy transfer per particle is larger. Conversely, the effect is less pronounced in the dark matter component, where the larger number of particles can more effectively share the energy exchange, minimizing the impact per particle. Additionally, as expected, the artifact becomes more significant as the mass ratio deviates more from 1.

We find that this effect is very similar for our method (at number ratios of $1:1$, $4:1$, and $7:1$) and the traditional methods (classical, $4:1$ and $7:1$ grid loads), suggesting the problem is independent of whether glass or grid initial conditions are used. Therefore, employing an equal-mass initial condition, as our method allows, is crucial for mitigating this numerical artifact in two-component cosmological N-body simulations.

\subsection{Non-radiative hydrodynamical runs}\label{subsec:app_hydro}

We repeat our analysis using a set of non-radiative hydrodynamical runs, with the resulting $z=0$ power spectra shown in the bottom row of Figure~\ref{fig:application_pk}. Note that in the ratio subpanels for these runs, we now use our equal-mass hydrodynamical run as the reference, since the L100N512 simulation is dark-matter-only. We observe qualitatively similar behavior to the pure gravity runs described in Section~\ref{subsec:app_pure_grav}. In the unequal-mass runs, spurious heating again causes the component with lighter particles to have its small-scale power suppressed, while the component with more massive particles to have its power enhanced. The effect is more pronounced in the gas component which contains fewer particles.

These results demonstrate that spurious collisional heating leads to noticeable numerical artifacts in unequal-mass simulations even when hydrodynamics is included, affecting the small-scale gas distribution in these non-radiative runs. This confirms that an equal-mass initial particle setup is crucial for two-component hydrodynamical simulations. A natural next step would be to investigate how these artifacts behave in galaxy formation simulations that include more complicated physical processes, such as radiative cooling, star formation, and feedback from stars and black holes \citep[see e.g.,][]{Ludlow2019,Ludlow2023}. While such an investigation is beyond the scope of this paper, we plan to explore this in future work using our particle loading method and the galaxy formation model described in \citet{Liao2023}.

\section{Summary} \label{sec:sum}

In this paper, we present a method for generating two-component glass-based particle loads that can achieve arbitrary number ratio, $N_{\rm DM} : N_{\rm gas}$, by extending the traditional glass-making approach. Specifically, we simultaneously relax two Poisson particle distributions under two anti-gravity forces: one from all other particles and the other from particles of the same component (Equations~\ref{eq:gas_force} and \ref{eq:dm_force}). We demonstrate that the generated particle load closely follows the expected minimal power spectrum, $P(k) \propto k^{4}$ (Figure~\ref{fig:pk}), exhibits good homogeneity (Figure~\ref{fig:uniformity}) and excellent isotropy (Figure~\ref{fig:isotropy}), and is sufficiently stable under gravitational interactions (Figure~\ref{fig:force_balance}).

We apply our method to two-component cosmological simulations by creating particle loads with different particle number ratios (and thus achieving different particle mass ratios). In both pure gravity and non-radiative hydrodynamical simulations, we find that due to spurious collisional heating, the component with lighter particles has its small-scale power suppressed, while the component with more massive particles has its power enhanced. Our results confirm that an equal-mass particle setup is crucial for mitigating this numerical artifact in cosmological simulations (Figure~\ref{fig:application_pk}).

We have implemented this method into the \textsc{gadget-2} code, which is publicly available at \href{https://github.com/liaoshong/gadget-2glass}{https://github.com/liaoshong/gadget-2glass}. It can generate particle loads for simulations that require equal-mass dark matter and baryonic particles under different cosmologies. Our approach can be readily extended to generate multi-component (i.e., more than two) uniform and isotropic particle distributions.

\begin{acknowledgments}
We are grateful to the anonymous referee for insightful comments that significantly improved the manuscript. We also thank Simon D. M. White for helpful discussions and comments. We acknowledge the supports by the National Natural Science Foundation of China (NSFC) grant (Nos 12588202, 12473015).
\end{acknowledgments}

\software{\textsc{gadget-2} \citep{Springel2005}, \textsc{matplotlib} \citep{Hunter2007}, \textsc{numpy} \citep{Harris2020}, \textsc{scipy} \citep{Virtanen2020}.}

\appendix

\section{Effects of the additional force terms} \label{ap:force_term}

To assess the effects of the additional force terms in Equations (\ref{eq:gas_force}) and (\ref{eq:dm_force}), we perform a series of test runs with different values for the factors $C_{\rm gas}$ and $C_{\rm DM}$. Note that, to isolate the impact of the extra evolution steps after switching off the additional force terms (studied in detail in Appendix~\ref{ap:steps}), all test runs presented in this appendix were evolved for a total of 8192 steps without switching off the additional force terms. All runs use particle numbers of $N_{\rm gas} = 64^3$ and $N_{\rm DM} = 1402203$.

\begin{figure*} 
\centering\includegraphics[width=\textwidth]{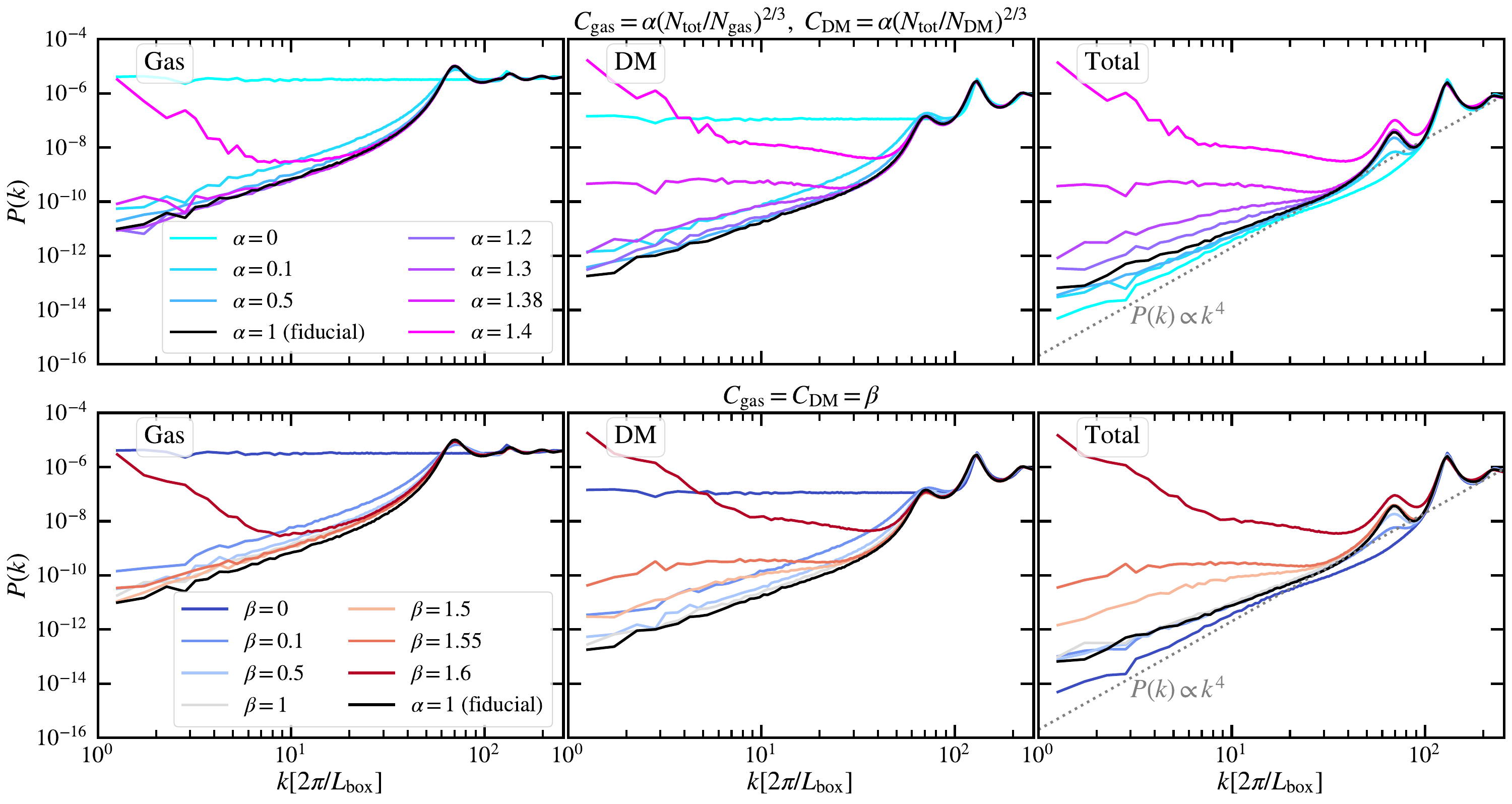}
\caption{Effects of the additional force terms in Equations (\ref{eq:gas_force}) and (\ref{eq:dm_force}) on power spectra. Top panels show the power spectra from test runs with varying $C_{\rm gas} = \alpha (N_{\rm tot} / N_{\rm gas})^{2/3}$ and $C_{\rm DM} = \alpha (N_{\rm tot} / N_{\rm DM})^{2/3}$, scaled to the fiducial values. Bottom panels show the power spectra from tests with different $C_{\rm gas} = C_{\rm DM} = \beta$. From left to right, the gas, dark matter, and total power spectra are plotted. The dotted line in the right panel shows the minimal power spectrum, $P(k) \propto k^{4}$. Deviating from the fiducial additional force terms leads to power spectra that depart from the fiducial curves (black). The fiducial choice, which is based on scaling arguments, is optimal in the sense that it yields the least large-scale power in gas and dark matter.}
\label{fig:fgas_fdm_test}
\end{figure*}

For the first set of test runs, we adopt $C_{\rm gas} = \alpha (N_{\rm tot} / N_{\rm gas})^{2/3}$ and $C_{\rm DM} = \alpha (N_{\rm tot} / N_{\rm DM})^{2/3}$, scaled to the fiducial values ($\alpha = 1$), with $\alpha$ varied from 0 to 2. As explained in Section~\ref{subsec:met_details}, the fiducial scaling is motivated by ensuring that the forces from the gas-only or dark matter-only particle systems are comparable to those from the full particle set. In the top panels of Figure~\ref{fig:fgas_fdm_test}, we plot the gas, dark matter, and total power spectra from a subset of these test runs.

With no additional force terms ($\alpha = 0$), the total particle system relaxes to a traditional glass distribution, and the total power spectrum follows the minimal $P(k) \propto k^{4}$ at large scales. In contrast, the large-scale power spectra of the gas and dark matter are dominated by Poisson noise, indicating that both components resemble uniformly random distributions. Even with $\alpha = 0.1$, the situation already improves significantly, i.e., large-scale power in both gas and dark matter is noticeably reduced, indicating that the additional force terms push the same-type particles apart. For $\alpha > 1$, the additional force terms dominate, the particle load relaxes to states that deviate from a glass, and the large-scale power spectra increase. Overall, the fiducial choice of $C_{\rm gas}$ and $C_{\rm DM}$ yields the lowest large-scale power in the gas and dark matter components while keeping the total power spectrum close to that of a traditional glass.

We also run a second set of tests with the simple choice, $C_{\rm gas} = C_{\rm DM} = \beta$, and the resulting power spectra are shown in the bottom panels of Figure~\ref{fig:fgas_fdm_test}. When $\beta$ deviates from $1$, the large-scale power in the gas and dark matter components increases, with $\beta = 1$ yielding the lowest large-scale power. However, compared to the $\beta = 1$ run,  the fiducial choice ($\alpha = 1$) still exhibits lower large-scale power in both gas and dark matter components, suggesting that it is preferable.

From our tests, we find that our fiducial choice for the additional force terms, based on scaling arguments, is practically optimal.

\section{Effects of the extra evolution steps} \label{ap:steps}

\begin{figure} 
\centering\includegraphics[width=0.825\textwidth]{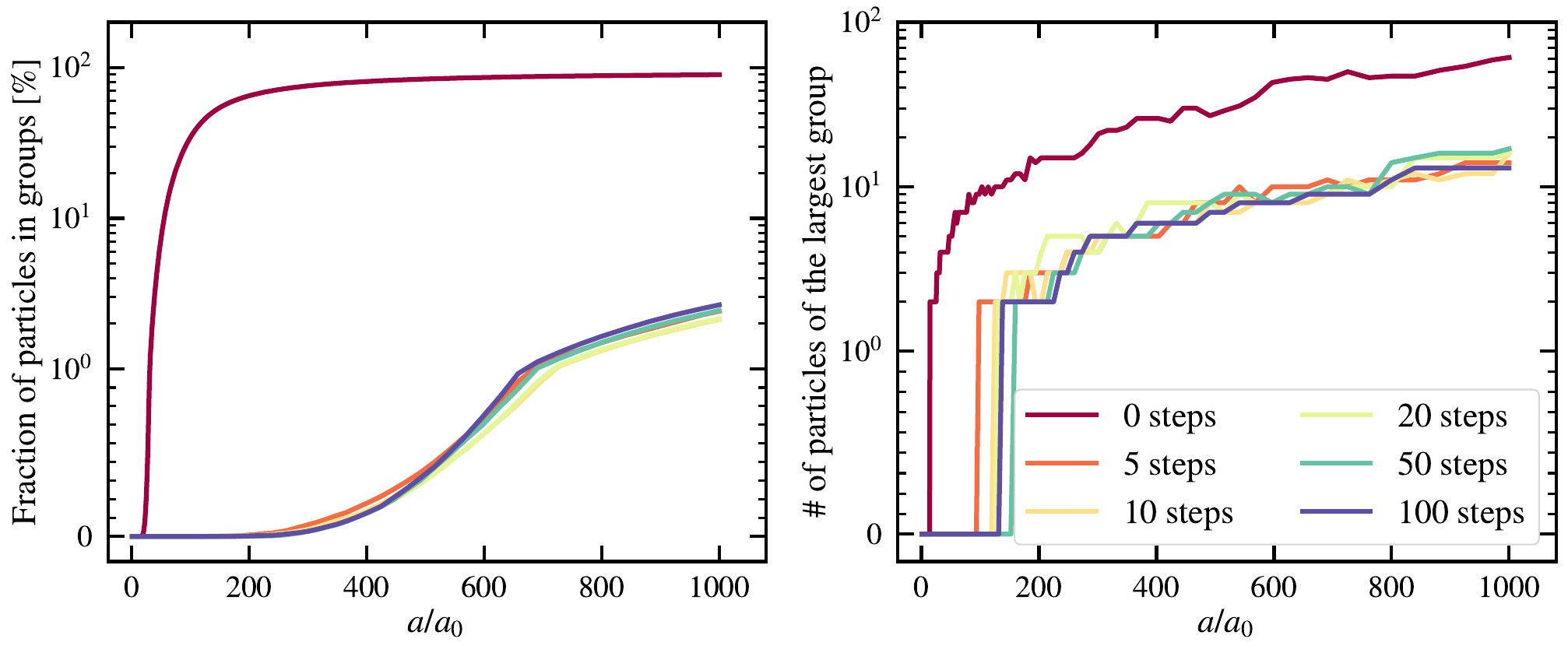}
\caption{Effects of performing extra evolution steps after switching off the additional force terms on force balance. Similar to Figure~\ref{fig:force_balance}, the left panel shows the time-evolution of the fraction of particles in the identified FOF groups, whereas the right panel shows the time-evolution of the number of particles for the largest FOF group. For clarity, the $y$-axis uses a linear scale in range $[0, 1)$ and a logarithmic scale for $y \geq 1$ in both panels. Without performing extra steps after switching off the additional force terms (the dark red line), structures start to form quickly (i.e., at $a/a_0 \sim 15$). By performing only a few extra evolution steps, the property of force balance for the resulting particle load can be significantly improved.}
\label{fig:fof_step_num_test}
\end{figure}

\begin{figure*} 
\centering\includegraphics[width=\textwidth]{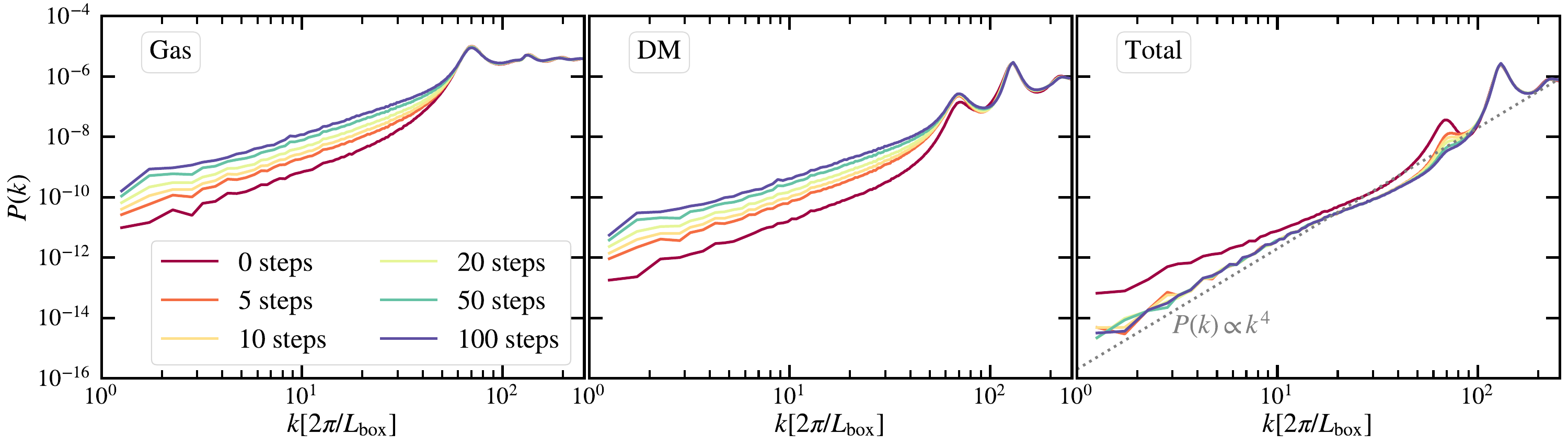}
\caption{Effects of extra evolution steps on power spectra. Different colors denote test runs with different numbers of extra steps. In the right panel, the dotted line indicates the minimal power spectrum, $P(k) \propto k^{4}$. An increased number of extra steps leads to an increase in large-scale power for the individual gas and dark matter components, while simultaneously decreasing it for the total particle set.}
\label{fig:step_num_test}
\end{figure*}

As stated in Section~\ref{subsec:met_details}, to improve overall force balance in the particle load, we evolve it for extra steps after switching off the additional force terms. To illustrate this and assess the effect of these extra steps, this appendix presents tests with varying numbers of extra evolution steps (0, 5, 10, 20, 50, and 100). We again set the particle numbers to $N_{\rm gas} = 64^3$ and $N_{\rm DM} = 1402203$ for these tests.

To study force balance, we evolve each final load from $a_0 = 0.001$ to $a = 1$ under pure gravity and identify FOF groups with at least 2 particles in all snapshots, as in Section~\ref{subsec:for_bal}. Figure~\ref{fig:fof_step_num_test} shows the time-evolution of the fraction of particles in the identified FOF groups and the number of particles for the largest FOF group. Without extra steps, the particle system quickly forms structures under gravity: the first two-particle group appears at $a /a_0 \sim 15$, and by $a/a_0 \sim 100$, the FOF particle fraction reaches ${\sim} 35 \%$, with the largest group containing ${\sim} 10$ particles. At the final time ($a/a_0 = 1000$), these values grow to $89.6 \%$ and $61$, respectively. In contrast, preparing the particle load with even a few extra steps (e.g., 5) significantly improves the force balance. In these cases, the final FOF particle fraction is $< 3 \%$, and the largest group has only ${\sim} 15$ particles.

However, we note that too many extra evolution steps can degrade the quality of individual components. Figure~\ref{fig:step_num_test} shows the power spectra for particle loads prepared with different numbers of extra steps. While evolving the system for extra steps brings the total power spectrum closer to the ideal $P(k) \propto k^{4}$ glass state, it also increases the large-scale power in the gas and dark matter components. This adverse effect is more pronounced with a larger number of extra steps.

Based on the tests shown above, as well as others with different $N_{\rm DM}:N_{\rm gas}$ ratios not presented here, we find that ${\sim} 20$ is a good choice for the number of extra evolution steps.

\section{Comparison with the previous method} \label{ap:compare}

In this appendix, we quantitatively compare the particle loads from our method with those from the \citet{Yoshida2003} method (i.e., combining two independent glasses). Since the \citet{Yoshida2003} method was proposed for a $1: 1$ particle number ratio, our first comparison uses equal numbers of gas and dark matter particles ($N_{\rm gas} = N_{\rm DM} = 64^3$). Our load is generated by evolving for $8192$ steps, with the additional force terms switched off for the last $20$ steps. For a fair comparison, the combined glass from the \citet{Yoshida2003} method is also evolved for $20$ steps under anti-gravity.

\begin{figure*} 
\centering\includegraphics[width=0.975\textwidth]{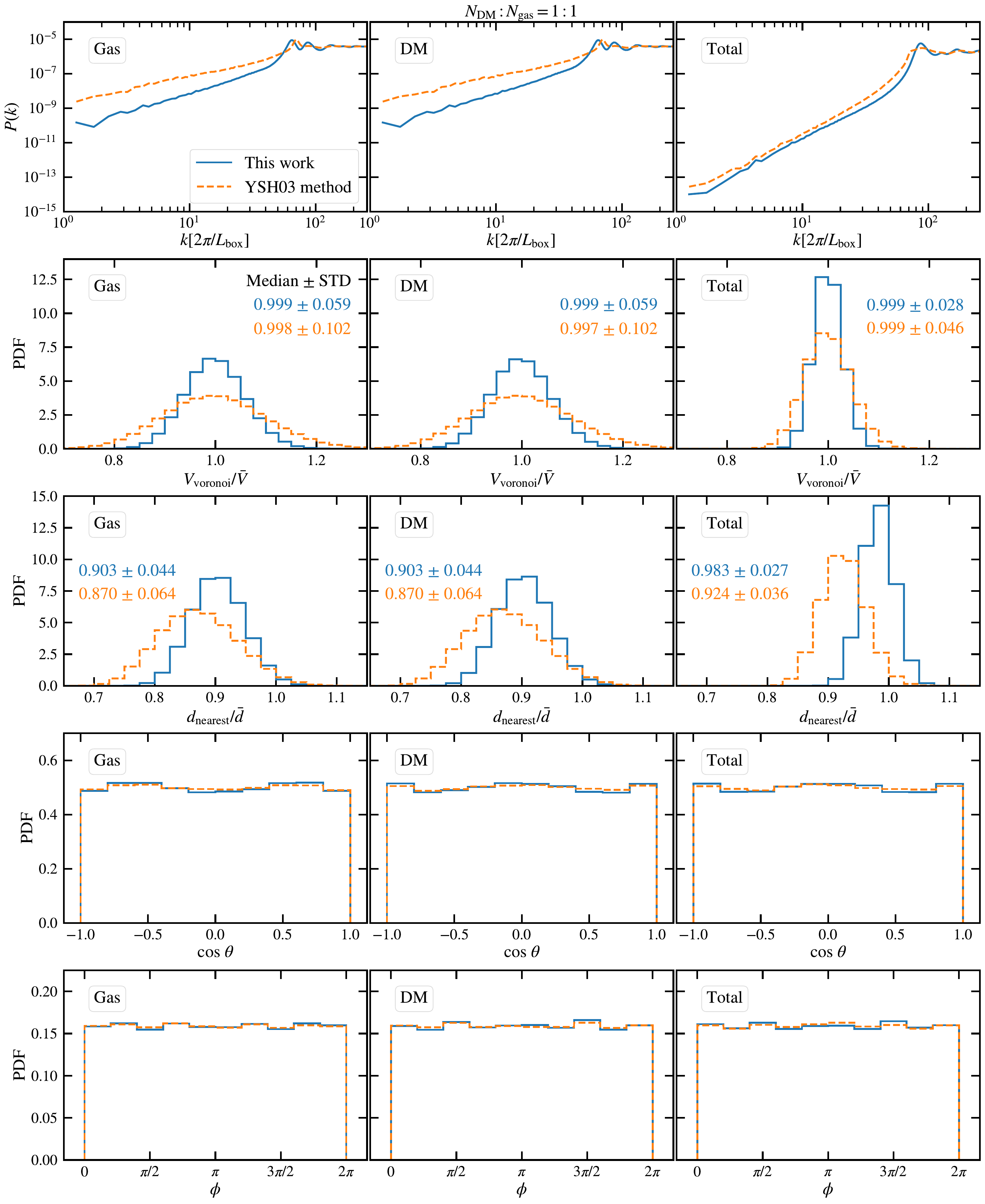}
\caption{Comparison between the method proposed in this work and that of \citet{Yoshida2003} for the case of $N_{\rm DM}:N_{\rm gas} = 1:1$. From top to bottom, we compare the power spectra, the distributions of Voronoi volumes, nearest-neighbor distances, polar angles, and azimuthal angles. The second and third rows show the median and standard deviation (STD) for each distribution (`Median $\pm$ STD'). From left to right, the results for the gas, dark matter, and total particle sets are plotted. Blue and orange colors denote our method and \citet{Yoshida2003}, respectively. Overall, relative to \citet{Yoshida2003}, our particle loads exhibit lower large-scale power and better homogeneity, while isotropy is similar for both methods.}
\label{fig:compare_previous_equal}
\end{figure*}

Figure~\ref{fig:compare_previous_equal} compares the statistical properties of these two loads. Specifically, we compare their power spectra, Voronoi volumes, nearest-neighbor distances, and angle distributions (see Section~\ref{subsec:uni_iso} for detailed definitions). Our method produces a load with noticeably better homogeneity. The power spectrum is lower on large scales, particularly for the individual gas and dark matter components, where it is an order of magnitude lower than the \citet{Yoshida2003} method. Furthermore, the distributions of Voronoi volumes and nearest-neighbor distances are tighter in our load, with standard deviations that are smaller (by approximately up to a factor of two) and medians that are closer to the ideal value of $1$. These features all point to a more uniform particle distribution. Both methods generate loads with excellent isotropy. The distributions of polar and azimuthal angles are nearly identical and consistent with a uniform random distribution, indicating that both loads are highly isotropic.

\begin{figure*} 
\centering\includegraphics[width=0.975\textwidth]{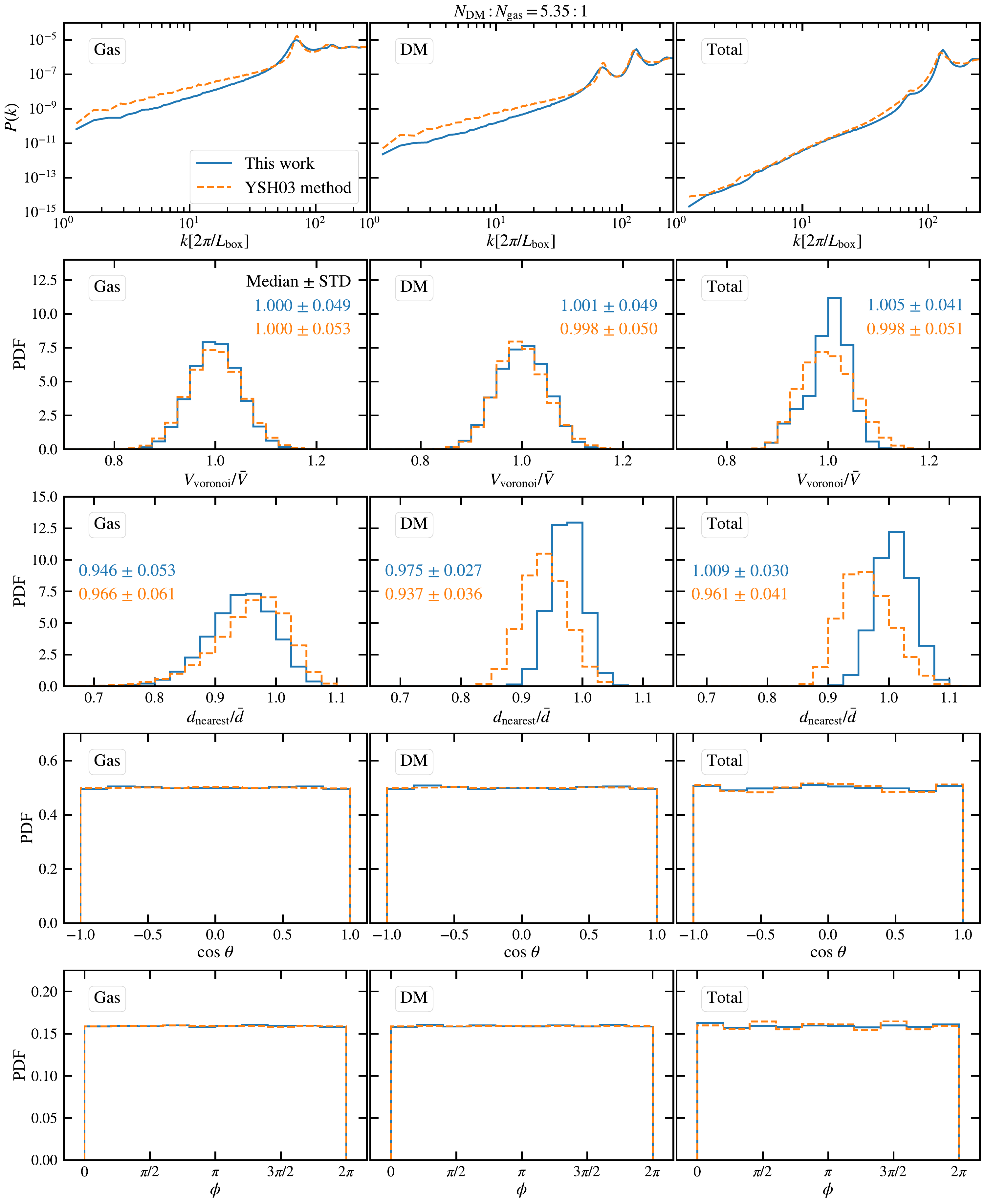}
\caption{Same as Figure~\ref{fig:compare_previous_equal}, but for the case of $N_{\rm DM} : N_{\rm gas} = 5.35 : 1$. Again, compared with \citet{Yoshida2003}, our particle loads show lower large-scale power (especially in the gas and dark matter components) and better homogeneity (notably in nearest-neighbor distances), while isotropy is similar for both methods.}
\label{fig:compare_previous_unequal}
\end{figure*}

We repeat the comparison for the unequal-number case, with $N_{\rm DM} : N_{\rm gas} = 5.35 : 1$ (specifically $N_{\rm gas} = 64^3$ and $N_{\rm DM} = 1402203$). The loads are prepared using a similar approach as before, and the results are shown in Figure~\ref{fig:compare_previous_unequal}. The differences between the two methods are similar to the equal-number case but less pronounced. We attribute this to the fact that the disruptive effects of the \citet{Yoshida2003} method are greater when particle numbers are more comparable (i.e., when $N_{\rm DM} : N_{\rm gas}$ is closer to $1$), as close juxtapositions affect a higher fraction of particles.

\bibliography{paper}{}
\bibliographystyle{aasjournal}

\end{document}